\documentclass[journal,transmag]{IEEEtran}
\usepackage{paralist,fancybox,listings, graphicx, algorithm2e,caption}
\ifCLASSINFOpdf
\else
\fi
\begin{document}
%
\title{Mitigating Data Exfiltration in SaaS Clouds}


\author{\IEEEauthorblockN{Duane Wilson, Ph.D.\IEEEauthorrefmark{1},
Jeff Avery\IEEEauthorrefmark{2}}
\IEEEauthorblockA{\IEEEauthorrefmark{1}Director of Strategic Cyber Initiatives,
Sabre Systems Inc, USA}
{\IEEEauthorrefmark{2}Department of Computer Science,
Purdue University, USA}
}

\markboth{Journal of \LaTeX\ Class Files,~Vol.~13, No.~9, September~2014}%
{Shell \MakeLowercase{\textit{et al.}}: Bare Demo of IEEEtran.cls for Journals}
%



\IEEEtitleabstractindextext{%
\begin{abstract}
Existing processes and methods for incident handling are geared towards infrastructures and operational models that will be increasingly outdated by cloud computing.  Research has shown that to adapt incident handling to cloud computing environments, cloud customers must establish clarity about their requirements on Cloud Service Providers (CSPs) for successful handling of incidents and contract CSPs accordingly. Secondly, CSPs must strive to support these requirements and mirror them in their Service Level Agreements.  Intrusion Detection Systems (IDS) have been used widely to detect malicious behaviors in network communication and hosts.  Facing new application scenarios in Cloud Computing, the IDS approaches yield several problems since the operator of the IDS should be the user, not the administrator of the Cloud infrastructure.  Cloud providers need to enable possibilities to deploy and configure IDS for the user - which poses its own challenges.  Current  research and commercial solutions primarily focus on protecting against Denial of Service attacks and attacks against the Cloud's virtual infrastructure.  To counter these challenges, we propose a capability that aims to both detect and prevent the potential of data exfiltration by using a novel deception-based methodology.  We also introduce a method of increasing the data protection level based on various threat conditions.
\end{abstract}

}

\maketitle

\IEEEdisplaynontitleabstractindextext

%
\IEEEpeerreviewmaketitle

\section{Introduction}
%
%
%
%
\IEEEPARstart{A}{ccording} to \cite{ref1}, security incident handling, an integral part of security management, treats detection and analysis of security incidents as well as the subsequent response activities (i.e., containment, eradication, and recovery) as outlined below:

\begin{itemize}
\item Detection: Discover indicators of possible security incidents
\item Analysis: Ascertain that indeed a security incident is at hand and understand what exactly has happened/is happening
\item Containment: Contain the incident before it spreads and overwhelms resources or increases damage
\item Eradication and Recovery Eliminate system changes caused by the incident and recover normal operations
\item Preparation and Continuous Improvement: Set up/adapt incident handling activities according to changing requirements and system/threat landscape.
\end{itemize}

Existing processes and methods for incident handling are geared towards infrastructures and operational models that will be increasingly outdated by cloud computing.  Research has shown that to adapt incident handling to cloud computing environments, cloud customers must establish clarity about their requirements on Cloud Service Providers (CSPs) for successful handling of incidents and contract CSPs accordingly. Secondly, CSPs must strive to support these requirements and mirror them in their Service Level Agreements. Typically, security incidents will cross the boundaries of both customer and CSP in each of the common deployment models (i.e., Software-as-a-service, Platform-as-a-service and Infrastructure-as-a-service or Saas, Paas, and IaaS respectively) denoting both joint responsibility and access. This must be taken into account when setting up incident handling or prevention for a cloud infrastructure.  Our research focuses on attacks confined to the SaaS deployment model due to the limited level of control a user has over the security of their data once it is stored within the cloud.\newline
\indent Intrusion Detection Systems (IDS) have been used widely to detect malicious behaviors in network communication and hosts. IDS management is an important capability for distributed IDS solutions, which makes it possible to integrate and handle different types of sensors or collect and synthesize alerts generated from multiple hosts located in the distributed environment.  Facing new application scenarios in Cloud Computing, the IDS approaches yield several problems since the operator of the IDS should be the user, not the administrator of the Cloud infrastructure. Extensibility, efficient management, and compatibility to virtualization-based context need to be introduced into many existing IDS implementations. Additionally, the Cloud providers need to enable possibilities to deploy and configure IDS for the user.  To date, a number of theoretical frameworks have been proposed to address these concerns. \cite{ref2, ref3, ref4, ref5, ref6, ref7,  ref8, ref9, ref10,
ref11, ref12,ref13,ref14,ref15,ref16,ref17}.  None, however, have been implemented in practice, hence, the concepts that are presented cannot be effectively validated.  Other research focuses on protecting the Virtual Machine (VM) within the cloud infrastructure from attacks - primarily focused on monitoring VM resources to maximize utilization or prevent Denial of Service attacks.   \cite{ref18,ref19,ref20,ref21,ref22,ref23,ref24, ref25, ref26,ref27}.\newline
\indent One major gap identified in our assessment of the current state of affairs in applying IDS concepts in a Public cloud setting is data exfiltration detection and prevention.  Data exfiltration is defined as the unauthorized transfer of data from a computer.  To address this particular gap, we propose a novel method for detecting and preventing data exfiltration using cyber deception \cite{deception}.  According to \cite{deception}, cyber deception is a \textit{deliberate act to conceal activity on a network, create uncertainty and confusion against an adversary's efforts to establish situational awareness and to influence and misdirect adversary perceptions and decision processes}.  To accomplish this, we focus on the Detection and Containment aspects of the incident handling process.  In the case of our research, containment refers to the efforts aimed at preventing the exfiltration from occurring - not preventing the spread of a cyber infection (e.g., worm or other malware).  Our approach is implemented such that cloud users do not have to rely on the protection mechanisms offered by the Cloud Storage Provider (CSP) which are often limited to common Data Loss Prevention techniques - which operate based on pre-defined policies.  The limitations of these approaches include: 1) can be bypassed by the use of encryption 2) are also under the control of the CSP (client-side or server-side) and 3) are not designed to prevent against unknown attacks (i.e., that are not specified in the deployed policy).  To address these and other limitations we propose the following contributions:
\begin{itemize}
\item A novel data anonymization technique using format-preserving encryption for producing deception file objects based on input data
\item The design of a novel capability to detect data exfiltration attempts in Software-as-a-Service (SaaS) cloud storage providers without having to rely on the provider or a middleware vendor using cyber deception.  
\item The design of a novel data protection mechanism that adjusts protection level based on threat conditions. 
\item A proposed prototype implementation (future work section) of our design across several popular cloud storage vendors to show its viability in a number of cloud settings
\end{itemize}

The presented approach is not designed to replace existing solutions, but provide a mechanism that increases the Defense in Depth of data within a SaaS cloud environment.  The rest of this paper is organized as follows:  Section 2 provides a background of the threat space we address with this work and presents related research, Section 3 describes the design of a proposed Cloud Storage Environment that is protected via Cyber Deception concepts, Sections 4 concludes the paper, and Section 5 provides areas for future work - primarily focusing on a prototype implementation across multiple Cloud Storage Providers. 

\section{Background}

As mentioned above, existing research largely has focused on the protection on Virtual Machines within Cloud infrastructures as well as on proposing frameworks that would aid in the detection/protection of data within the cloud. None of the aforementioned methods resulted in implementations that could be used by the public to protect data in SaaS storage infrastructures.  The following research outlines a variety of methods attempting to remedy this.  One common weakness with those discussed below is that the user is not in control of the protection mechanisms applied to their data.  Hence, they have to trust that 1) the protections offered are adequate, 2) the audit information is accurate, and 3) the protections will not fail in the event of adversarial compromise.

\subsection{Threat Model}

When user data is stored in a Software as a Service (SaaS) Public Cloud setting, it is subject to the security features provisioned by the CSP.  As mentioned above, this leads to a number of threats that user's data can be subject to without their knowledge or control.  We focus our efforts on the methods by which sensitive data can be exfiltrated once it has been transferred to the cloud.  Our research does not examine uploads for policy conformance.  We focus specifically on data that is resident on the cloud - as we assume that users have followed any relevant policy that governs the storage of data in the cloud.  Within the cloud, there are 3 primary ways data can be accessed and hence subsequently exfiltrated: 1) Via Sharing (individual or group), 2) Via download (web interface or client application), and 3) Via the cloud interface (within the cloud environment).  Our research focuses on the detecting and prevention of data exfiltration via sharing and download requests.  As previously stated, SaaS environments are not under the control of the user - hence we cannot modify the cloud infrastructure to monitor accesses within the environment itself.  An additional feature we provide is the ability to be aware of varying threat conditions - providing a higher level of protection when the adversarial threat is at its peak.     

Within the cloud setting, data exfiltration can be prevented via the following methods: 1) Detect an attempt and prevent it before it happens - would require the monitoring all potential paths of theft in cloud and intercepting malicious looking attempts, 2) Allow/Disallow information from being downloaded based on individual or organizational policies (after data is stored), or 3) Allow access to information in cloud only when given permission or authorization.  In the SaaS deployment model, method 1 is not possible without the modification of the Cloud infrastructure.  Method 2 focuses on protecting information from being exfiltrated via the Cloud - not after the information is stored.  This has also been addressed by CloudFilter (a paper we discuss below).  We aim to control data exfiltration via the 3rd method to prevent attempts by malicious adversaries who have compromised a user account \cite{ref29}.  We assume that users are generally trustworthy (i.e., not malicious insiders) and are able to make wise decisions about who needs to have access to their data - prior to it being uploaded.  Our approach can be used as an alternative or augment to client-side encryption cloud storage vendors that offer a method of protecting user data from the cloud provider themselves.  

\subsection{Related Work}
\subsubsection{Cyber Deception:}
The art of deception has been used as a technique to lure attackers away from important data and learn about their techniques and targets\cite{stoll,planning_deception}.  In the space of cloud data protection, two methods have been proposed: Fog Computing and a Honeypot implementation for the cloud.  In \cite{ref30}, the authors propose a novel deception-based concept that focuses on the protection of insider data theft attacks by monitoring data access in the cloud and detecting abnormal access patterns.  They accomplish this by profiling the normal behavior of users and upon the detection of an anomaly, they produce and return decoy information (e.g. docs, honey files, honeypots) to the attacker.  Being that there is no implementation in an actual cloud environment, it is difficult to ascertain the viability of their approach due to the fact that quantifying normal user behaviors is a difficult problem\cite{anomaly_detect_challenges} - in general - and much more in a cloud computing environment that is largely multi-tenant.  Other approaches \cite{ref31, ref32} attempt to apply anomaly-based detection mechanisms to the cloud but suffer from the same issues- profiling users in such a complex environment and not being implemented in an actual cloud environment.  

Similarly, \cite{ref33} aims to employ deception concepts in a private cloud using a honeypot.  The authors deploy the Snort Network Intrusion Detection System (NIDS) in the Eucalyptus private cloud as a proof of concept implementation.  They incorporate a honeypot into the NIDS to further entice attackers and provide a mechanism to learn more about similar future attacks.  The primary weaknesses of their approach is that 1) it would require a major modification to existing cloud infrastructures to implement their methodologies, 2) it uses a signature-based approach for detection, and 3) their deployment would require awareness of Cloud infrastructure to enable full OS coverage (e.g. Windows, Linux, OS X). Neither the Fog Computing, Honeypot, or Anomaly-Based approaches as described would be feasible in practice due to these limitations.  Our proposed solution extends the concepts present within Fog Computing by providing an implementation in an actual cloud storage environment, providing a novel mechanism for users to generate decoy information that is derived from their uploaded files, and focusing specifically on data exfiltration problem. Additionally we provide a mechanism that can adjust to current threat conditions as we subsequently discuss.

\indent \textbf{Commercial Cloud Providers: } There are a number of Commercial Cloud \newline Providers that provide features to protect the security of cloud user data.  The primary security features they all provide are: the protection of data in transit and data at rest and two-factor authentication.  Other than Box, they all offer limited oversight of the activities associated with files/folders.
Within Dropbox, the event log tracks when users create, delete, and restore Dropbox folders. Once an action is taken, the event log records the name and ID of the Dropbox folder, the action, the user who made the change, and the date it was performed \cite{ref34}.  For Google Drive, the types of file activities that are recorded are the: 1) moving and removing, 2) renaming, 3) uploading, 4) sharing and unsharing, and 5) editing and commenting. There are no specific additional security features provided in the documentation \cite{ref35}.  SugarSync\footnote{www.sugarsync.com} monitors any edits to a file, any new files created, files deleted, and changes to a folder or sub-folders \cite{ref36}.  Box \cite{ref37} offers the most advanced native security features of all the existing CSPs (to include the ones described above). Box provides comprehensive reporting, logging, and audit trails in order to track account activity, file access, settings changes and nearly everything else that occurs in Box. \newline \indent These features allow users to monitor for access violations only, but are not combined with a mechanism to prevent data loss.  Box, however, does provide a separate mechanisms to detect and prevent data loss via extended security policies that allows users to keep content confidential, mitigate data loss, flag risky sharing requests, and white list certain domains that should be trusted.  Some native features are present, but others are offered via collaborative security partners.  CipherCloud (one such partner) offers of a secure middleware service that provides users with insight into cloud data \cite{ref36}. They provide a range of features to include: 1) User activity monitoring, 2) Anomaly Detection, 3) Comprehensive Audit Logs, and 4) Ongoing Data Loss Prevention Monitoring.  This solution has a number of promising features, however, these all require trust of an entity like CipherCloud with their data (risk transference) and it also comes at a premium cost and is geared towards organizations - not individual users.

\textbf{CSP IDS Implementations: } In\cite{ref38}, Greg Roth and Don Bailey present a detailed overview of Amazon Web Services (AWS) Intrusion Detection capabilities.  These capabilities are a combination of several features: AWS's Identity and Access Management, Multi-Factor Authentication, Amazon S3 Bucket Logging, Security Audit Role, Write Once Storage for Data Provenance (Versioning), and Auditing Logs.  Their approach is to use this compendium of features to detect unauthorized access to user data based on user roles.  The major weakness of this role-based approach is that it is a script based IDS, not an actual system that the general user can use.  Another weakness is dynamically adjusting to changing roles may be difficult to capture and categorizing employee's using clear cut roles may be difficult to define.  Due to the number of components involved, it would require a security expert to setup, configure, and interpret the resulting alerts.  It is limited in scope - focusing primarily on access permission violations not specific attack types (e.g., data exfiltration, integrity breaches) and is specific to Amazon EC2, hence, not interoperable with other CSP infrastructures.  

Lastly, in \cite{ref39}, Ioannis and Pietzuch describe CloudFilter - a system for the practical control of sensitive data propagation to the cloud.  Their scheme monitors HTTP File Transfers to the Cloud independent of the CSP being used.  Upon upload of each user file or files, CloudFilter enforces data propagation policies to confirm that the upload request is authorized.  Their approach is limited to just monitoring the HTTP protocol which would cause CloudFilter to miss sensitive data being propagating via another protocol.  One major strength of the described approach is that no modification is necessary to the CSP infrastructure, however, it does require modification of files upon upload and web browsers for user identification.  This solution does not address data exfiltration once it is resident within the cloud, because it only monitors uploads.  

\section{Prototype Design}
\begin{figure}[h]
\centering
  \includegraphics[width=9cm,height=4.5cm]{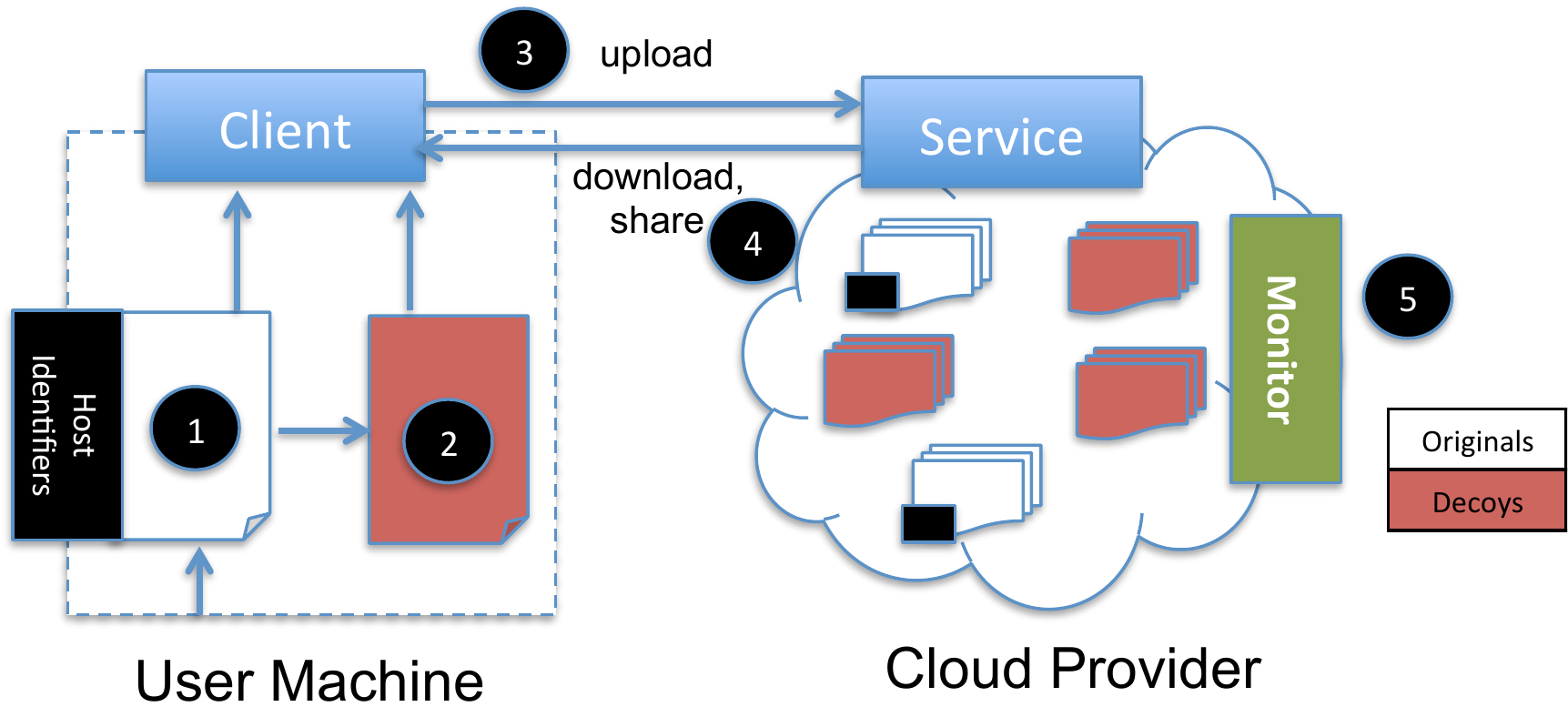}
  \caption{Proposed Prototype Architecture: Prior to entering deceptive file editing, INFOCON Level is set. 1) A file is submitted and host system identifiers are embedded in the file 2) Files are deceptively modified 3) Upload and store original and decoy document to cloud 4) A file within the cloud is access  5) Monitor exfil attempt(s) for a particular file }
  \label{fig:flow}
\end{figure}

The goal of our prototype is to provide a transparent and practical solution for protecting sensitive data stored in Software-as-a-Service (SaaS) Cloud Storage Provider, which represent stored data as files. An important requirement is for the system to be applicable across different cloud storage providers with minimal end-user configuration.  We accomplish this by leveraging a cloud agnostic API (discussed in next section) to avoid having to adapt to the API of specific cloud storage service. Secondly, the deceptive information should be automatically generated based on the input provided to accompany the files that are uploaded.  Lastly, we provide a solution that can be employed in a representative scenario based on the criticality of the cyber situation and/or INFOCON status.  We accomplish this without modification of the cloud infrastructure. Our prototype provides the following component engines all integrated with a cloud storage provider:
\begin{itemize}
\item \textbf{Generation engine} that produces deceptive Objects or Decoy Documents with high similarity to original content
\item \textbf{Detection engine} that monitors access requests and verifies originating source. 
\item \textbf{Prevention engine} that controls access to legitimate documents via deception-based cloud access layer 
\item \textbf{Threat engine} that makes adjustment of deception-based protection ``level'' based on varying threat conditions (e.g., INFOCON Level)
\end{itemize}
Below we walk through a exfiltration attempt against our framework. There are several steps that take place between a user (or enterprise) and a cloud provider during an upload operation.  Prior to step 1, the initial threat level is set to INFOCON Level 5 - indicating the lowest threat environment.  In step 1, a user submits a file via our prototype application and several host system identifiers are embedded within the file to include: Computer MAC Address, Network IP Address, Network Hostname, UserID, and Hash of 4-tuple identifiers (MAC, IP, Hostname, UserID). In step 2, the file is read and each numerical data element of the file is anonymized - creating the decoy document.  Sensitive data within a file is usually represented as numerical data (e.g., SSN, Credit Card), but can also be represented as names such as locations and people.  Numerical data can be recognized using regular expressions.  Name data can be recognized using Natural Language Processing techniques such as Named Entity Recognition\cite{ner}.  Once identified, sensitive data can be changed to a random value or to a more strategic value using look up tables and online searches depending on the level of security necessary. Step 3 results in both the decoy document and the original document being uploaded and stored in the cloud.  Steps 1-3 represent how a file is uploaded to the cloud. Once the file is uploaded, it persists on the cloud as any other file uploaded.  Next, assume some time as passed since the files were uploaded to the cloud.  Step 4 represents a request(s) for a particular file stored within the cloud.  Lastly, step 5 commences the monitoring process for exfiltration attempts and threat level updates.  Steps 4-5 represent an exfiltration attempt on some file. Figure \ref{fig:flow} provides an illustration of our prototype architecture.

A few foundational concepts in cyber deception have preceded our design and subsequent implementation - which we outline here for context.  Honeyfiles are bait files intended for hackers to access. The files reside on a file server, and the server sends an alarm when a honeyfile is accessed \cite{ref41}.  In  \cite{ref42}, Wang et. al. add honey activity to enhance the realism of honeyfiles. The Decoy Document Distributor (D3) System is a tool for generating and monitoring decoys.  An accompanying website (FOG) allows users to download files, such as tax documents and receipts, that appear authentic but actually contain spurious information \cite{ref43}. Lastly, Fox provides a mechanism for placing decoy documents in places that are more likely to get accessed \cite{ref44}. As previously stated, there has been much previous work in the area of  using  deceptive  data  to  defend against malicious adversaries. Yet no prior work has attempted to adapt these concepts in a SaaS cloud storage environment - which is out of the control of a user.  Secondly, no work has been proposed that provides the ability to adjust to varying threat situations (e.g. INFOCON levels).  Additionally, we address the specific threat of data exfiltration in SaaS storage environments.  Lastly, our approach uses pre-existing content to generate deception data whereas previous work focuses on generating data based on sensitive properties.  

For decoy object generation, we follow a different process than is used to generate a honeyfile, where users are responsible for selecting the honeyfiles in their environment.  In our scheme, we rely on data within each file to generate a replica highly similar to original content.  Creating similar documents can influence an attacker's decision making.  Not knowing which document is the original and therefore what content is legitimate  can force them to make incorrect decisions and potentially critical errors.  Creating a higher ratio of replica to original files makes the this technique probabilistic in that the ability of an attacker to guess the correct file if inversely related to the number of replica files.  For the generation process, each file is tokenized (primary delimiter would be spaces) into individual words and the word structure examined to determine its sensitivity.  We make a simple assumption that sensitive content will likely take a numerical format (not including alphanumeric characters).  Hence, if a word is determined to be sensitive it is converted using format-preserving encryption.  In cryptography, format-preserving encryption (FPE) refers to encrypting in such a way that the output (the ciphertext) is in the same format as the input (the plaintext) \cite{ref45}. Upon completion of the document analysis process, the decoy document will be created.  The naming convention for the decoy document will be similar to the original to ensure plausibility.  

As previously stated, there are several ways that data can be exfiltrated via the cloud: Downloading, Sharing, or Cloud Access.  To adequately detect and prevent exfiltration through these channels, we provide a system that monitors all requests to information residing on a cloud storage provider's infrastructure from a host system.  We are unable to monitor access requests originating from the Cloud Provider itself to the design of SaaS environments.  Hence, we focus on Download and Sharing operations.  Whenever a download or sharing request is made, the default identifier (e.g. host MAC address) is used to confirm the legitimacy of the request.  We use the MAC address because it is unique to each machine.  Though it can be spoofed, it is a metric that we have access to that is unique to each machine\footnote{Ideally, a value that is unique to every machine and stored in a secure location on the device (e.g. a value created and stored on the Trusted Platform Module\cite{tpm}) should be used.}.  If the originating host identifier matches the embedded identifier within the file, the actual file is returned to requesting party.  Otherwise the corresponding decoy document is returned and the event is logged. Increasing levels of protection are imposed as the threat level increases as will be discussed below.   

Cyber threats commonly change over time within a particular environment.  We argue that this is also the case within a cloud storage environment.  Hence, a solution presented to protect data within that environment, should be able to adapt to the ebb and flow of the threat landscape.  Hence our approach proposes an emulated threat environment that mirrors the INFOCON levels \footnote{Strategic Command Directive (SD) 527-1 (2006-01-27). "Department of Defense (DOD) Information Operations Condition (INFOCON) System Procedures" (PDF). DISA Policy and Guidance. Retrieved 2016-04-03.} used by the US government to designate the presence of a hostile threat.  With each threat level increase, our approach requires additional system identifiers to validate an access request.  We describe the U.S. government INFOCON levels below: 
\begin{itemize}
\item INFOCON 5 describes a situation where there is no apparent hostile activity against computer networks. Operational performance of all information systems is monitored, and password systems are used as a layer of protection.
\item INFOCON 4 describes an increased risk of attack. Increased monitoring of all network
activities is mandated, and all Department of Defense (DoD) end users must make sure their
systems are secure. Internet usage may be restricted to government sites only, and backing up files to removable media is ideal. 
\item INFOCON 3 describes when a risk has been identified. Security review on important systems
is a priority, and the Computer Network Defense system's alertness is increased. All
unclassified dial-up connections are disconnected.
\item INFOCON 2 describes when an attack has taken place but the Computer Network Defense
system is not at its highest alertness. Non-essential networks may be taken offline, and alternate methods of communication may be implemented. 
\item INFOCON 1 describes when attacks are taking place and the Computer Network Defense
system is at maximum alertness. Any compromised systems are isolated from the rest of the
network.
\end{itemize}

\section{Conclusions}
In this paper, we have proposed several methods to address the problem of data exfiltration detection and prevention in a SaaS Cloud Storage environment.  Our proposed methods will overcome the limitations of existing approaches that primarily focus on the protection of the virtual infrastructure - not the data that is stored in these environments.  We leverage a number of concepts in cyber deception for data protection and introduce the ability to adapt to varying threat conditions.  
Data Loss Prevention (DLP) is often implemented within a cloud storage environment to protect against data exfiltration.  However, we 
found that DLP has a number of shortfalls to include: 1) can be bypassed by the use of encryption 2) are also under the control of the CSP and 3) are not designed to prevent against unknown attacks. As a result, we demonstrate the added value of using decoy documents within a SaaS environment for more dynamic data protection.  We provide a prototype implementation that interoperates with a number of popular cloud storage providers and offer supporting analysis results.  

\section{Future Work}
In future work, we will provide a prototype implementation of the design in section 3.  The goal of our prototype implementation will be to demonstrate the viability of our solution across a number of different Cloud Storage Providers (CSPs).  We will present implementations for Dropbox\footnote{www.dropbox.com}, Google Drive\footnote{www.google.com/drive/}, and Box\footnote{www.box.com} CSPs.  We will leverage a number of 3rd party APIs to enable the key features for our prototype: kloudless, Extensible Metadata Platform (XMP) toolkit, and libffx.  For each CSP, the implementation focus will be on 1) Generation of Decoy Documents, 2) Detection and Prevention of Data Exfiltration Attacks, 3) Adjusting the Protection Levels Based on Threat Conditions.

Addtionally, we plan to integrate several threat database sources to provide higher fidelity information prior to making a data protection adjustment (e.g. threatconnect, threatscape, enigmadatabase).  Additionally, we will identify other actions that could be taken upon notification of an increased threat level (e.g., increase the ratio of decoy objects in the environment or impose a data access timeout period).  As a proof of concept, we focused on the action that was most effective in defending against data exfiltration attempts.  Addressing other stages within the cyber kill chain model would be a natural evolution of our concepts.  The goal would be to address different aspects of the user experience in the cloud (from a security perspective) with various cyber deception artifacts.  In this work we have implicitly focused on mitigating the `action on objectives' stage of the kill chain process - the last phase of the kill chain.  Future work would seek to address the Reconnaissance, Weaponization, Delivery, Exploitation, Installation, and Command \& Control phases.  


%

\ifCLASSOPTIONcaptionsoff
  \newpage
\fi

\end{document}